\documentclass[11pt,a4paper]{article}
\pdfoutput=1
\usepackage{jheppub}
\usepackage{cite}

\usepackage{multirow, graphicx,amssymb,url,mathrsfs,amsmath}
\usepackage{wrapfig,boxedminipage,subfigure,epsfig}
\usepackage{amsxtra,amstext,latexsym,dsfont,amsfonts}
\usepackage{color}
\usepackage[dvipsnames]{xcolor}
\usepackage{float}
\usepackage{slashed}
\usepackage{calligra}
\usepackage{enumerate}
\usepackage{comment}
\DeclareFontShape{T1}{calligra}{m}{n}{<->s*[2.2]callig15}{}
\DeclareMathAlphabet{\mathcalligra}{T1}{calligra}{m}{n}

\newcommand{\be}{\begin{equation}}
\newcommand{\ee}{\end{equation}}
\newcommand{\bea}{\begin{eqnarray}}
\newcommand{\eea}{\end{eqnarray}}

\title{Construction of bulk solutions for towers of pole-skipping points}

\author[]{Keun-Young Kim,}
\author[]{Kyung-Sun Lee,}
\author[]{and Mitsuhiro Nishida}

\affiliation[]{School of Physics and Chemistry, Gwangju Institute of Science and Technology, 123 Cheomdan-gwagiro, Gwangju 61005, Korea}

\emailAdd{fortoe@gist.ac.kr}
\emailAdd{kyungsun.cogito.lee@gmail.com}
\emailAdd{mnishida@gist.ac.kr}

\abstract{
The pole-skipping phenomenon has been proposed as a connection between chaotic properties of black hole geometries and special points where regular solutions of linearized Einstein equations at horizons have extra free parameters. In this work, we pursue the special points in the near-horizon analysis of integer spin-$\ell$ fields on the Rindler-AdS black hole. We construct linear combinations of field components to simplify coupled equations of massive fields and investigate towers of the special points along with imaginary Matsubara frequencies $i\omega=2\pi(n+1-\ell)T$ with a non-negative integer $n$ and the Hawking temperature $T$. We also propose that integrals of spin-$\ell$ bulk propagators over horizons of static black holes capture behaviors at the special points, which are generalizations of integrals of graviton propagators for shock wave geometries. Their interpretation is provided in terms of four-point amplitudes with the spin-$\ell$ exchange. 
 }

\begin{document}
\maketitle



\section{Introduction}
In quantum field theories, correlation functions are essential quantities to compute correlations between different points and scattering processes of fundamental particles. Depending on the application, the correlation functions in position space and momentum space are used, which are connected by a Fourier transformation.
For example, Feynman rules can be formulated in position space and momentum space.

As a recent development, out-of-time-order correlation functions (OTOCs)\cite{larkin1969quasiclassical,Kitaev-2014}, which are correlation functions in position space, have been proposed as a measure of quantum chaos in large $N$ quantum field theories. Specifically, Lyapunov exponent $\lambda_L$ and butterfly velocity $v_B$, which are determined from exponential behaviors of sub-leading terms in the large $N$ expansion of four-point OTOCs, diagnose quantum chaos. It is also conjectured that the Lyapunov exponent of consistent quantum field theories is bounded by the exponent in the theories that have the Einstein gravity duals \cite{Maldacena:2015waa}.

It has been studied that certain points called pole-skipping points are related to the chaotic properties of the theories with the Einstein gravity duals. These pole-skipping points in momentum space are points such that values of momentum Green's functions are not uniquely determined as functions of several complex variables. More explicitly, the exponents $\lambda_L$ and $v_B$ in the theories with the Einstein gravity duals are related to pole-skipping points of retarded two-point Green's functions of energy density. This relation is called a pole-skipping phenomenon \cite{Grozdanov:2017ajz, Blake:2017ris}.

If we assume the holographic dictionary, pole-skipping points can be detected by near-horizon analysis on the bulk side \cite{Blake:2018leo}. This near-horizon analysis searches conditions that make incoming regular boundary conditions on a black hole horizon non-unique, and we call points that satisfy such conditions special points.  Non-uniqueness of the incoming regular boundary conditions at special points can make retarded Green's functions on the boundary theory side ambiguous due to the holographic dictionary. One can further find special points in bulk equations of general fields on various backgrounds. See \cite{Haehl:2018izb,Grozdanov:2018kkt,Grozdanov:2019uhi,Blake:2019otz,Li:2019bgc,Natsuume:2019sfp,Natsuume:2019xcy,Jensen:2019cmr,Ahn:2019rnq,Das:2019tga,Haehl:2019eae,Natsuume:2019vcv,Balm:2019dxk,Wu:2019esr,Ceplak:2019ymw,Abbasi:2019rhy,Liu:2020rrn,Liu:2020yaf,Ahn:2020bks,Abbasi:2020ykq,Jansen:2020hfd,Grozdanov:2020koi,Ramirez:2020qer,Choi:2020tdj,Ahn:2020baf,Natsuume:2020snz,Arean:2020eus,Kim:2020url,Sil:2020jhr,Yuan:2020fvv,Abbasi:2020xli,Ceplak:2021efc,Baggioli:2021xuv,Wu:2021mkk,Blake:2021wqj,Jeong:2021zhz,Jeong:2021zsv, Kim:2021hqy,Natsuume:2021fhn, Yuan:2021ets, Blake:2021hjj} for recent studies of pole-skipping points and near-horizon analysis. 

The pole-skipping points of fields other than the energy-momentum tensor do not seem to be related to the maximally quantum chaos because the maximally quantum chaos will arise from the dominant contribution of the energy-momentum tensor exchange in four-point OTOCs. As a generalization of the original pole-skipping phenomenon, one can expect that the pole-skipping points of scalar and vector fields are related to the scalar and vector exchange in the four-point OTOCs. The authors of \cite{Ahn:2020bks, Kim:2020url} showed this expectation in terms of conformal field theories (CFTs) and holography. Moreover, in our paper \cite{Kim:2021hqy}, we also showed relations between the pole-skipping points in CFTs on the Rindler spacetime and the Regge limit of conformal blocks with the exchange of any integer spin by using the Rindler-AdS black hole. In order to understand the structure of generalized pole-skipping phenomena like these studies, researches on the pole-skipping points of various fields are useful. This is a motivation to study the pole-skipping points of fields other than the energy-momentum tensor.

Another motivation is to understand the pole-skipping phenomenon in the string theory. The string theory contains an infinite number of higher spin fields. When we consider gravity theories as a low energy limit of the string theory, we can ignore the higher spin fields,  and the maximally quantum chaos in the gravity theories is controlled by spin-two gravitons. If we would like to study the pole-skipping phenomenon in the string theory, we should consider the pole-skipping points of higher spin fields. Although there are subtleties due to the infinite number of fields, researches on the pole-skipping points of higher spin fields will be a good starting point for studying the pole-skipping phenomenon in the string theory.

For bulk fields with spin-$\ell$, it has been observed that there are towers of special points at frequencies $i\omega=2\pi (n+1-\ell)T$ (see, for instance, \cite{Grozdanov:2019uhi,Blake:2019otz,Natsuume:2019vcv,Wu:2019esr,Ceplak:2019ymw,Ahn:2020baf}), where $n\in\{0\}\cup\mathbb{N}$, and $T$ is the Hawking temperature of a static black hole. It is important to check whether the towers of special points exist in massive fields with $\ell\ge1$. For the massive fields, we cannot use gauge-invariant combinations for massless fields. Thus, we need to analyze coupled bulk equations or find linear combinations of field components for simple analysis.

The pole-skipping phenomenon implies that the leading special points of gravitons with $n=0$ are related to shock waves on black hole horizons that lead to $\lambda_L$ and $v_B$ in the holographic theories. However, even for the energy-momentum tensor, we do not have a good interpretation of the sub-leading special points. For a further interpretation of the sub-leading special points, it is significant to generalize a construction of the shock waves for the sub-leading pole-skipping points of various fields with $n\in\mathbb{N}$.

In this paper, we analyze the overall structure of special points of free fields with integer spin on the Rindler-AdS black hole. Along with the well-analyzed tower structure of scalar fields ($\ell=0$) \cite{Grozdanov:2019uhi,Blake:2019otz}, the special points of vector fields ($\ell=1$) and rank-$\ell$ higher-spin fields are investigated. So far, the tower structure of special points with $\ell=1,2$ in the near-horizon analysis has been studied only in massless cases by using a gauge-invariant method \cite{Blake:2019otz}.\footnote{The leading special points of  massive fields in the near-horizon analysis have been examined. See, for example, \cite{Liu:2020yaf, Kim:2020url}.} This is because one can define gauge-invariant variables for massless fields to decouple the equations of motions (EOMs) of vector and tensor fields, which are complicatedly coupled differential equations. In this study, we do not restrict ourselves to the analysis of the massless fields and do not use the gauge-invariant variables to decouple the EOMs. Rather, we use specific linear combinations of the fields to decouple the EOMs in a straightforward way. We obtain the special point structure for $\ell=0,1,2,3$ and induce results for the general spin-$\ell$ fields.

Assuming the existence of special points at $i\omega=2\pi (n+1-\ell)T$, we propose integrals of bulk propagators in position space over black hole horizons that capture behaviors at the special points of bosonic fields. Our construction of the integrals can be applied to the propagators satisfying a property about isometries of static black holes. We also discuss the interpretation of our construction from the viewpoint of tree-level scattering in the bulk. 

Our analysis suggests a connection between the sub-leading pole-skipping points with spin-$\ell$ and bulk integrals of spin-$\ell$ propagators, which are generalizations of the connection between the leading pole-skipping points of the energy-momentum tensor and bulk integrals of graviton propagators for shock wave geometries. This connection gives a generalized picture of the pole-skipping phenomena with various spin beyond the pole-skipping phenomenon for the maximally quantum chaos.

The paper is organized as follows. In Section \ref{sec:nha}, we study near-horizon analysis of massive fields with integer spin on the Rindler-AdS black hole. We construct integrals of bulk propagators that capture behaviors at special points of scalar fields in Section \ref{sec:si} and of spin-$\ell$ fields in Section \ref{sec:sli}. In Section \ref{sec:interpret}, we discuss the interpretation of the integrals. We conclude in Section \ref{sec:con} with a discussion.

\section{Near-horizon analysis of bulk fields with integer spin-\texorpdfstring{$\ell$}{l} on the Rindler-AdS black hole}\label{sec:nha}

In this section, we search for special points of spin-$0,1,\ell\,(\ell\geq2)$ fields, not only the leading special points but also the sub-leading special points. To begin with, we briefly review near-horizon analysis to detect the special points. Here we study on the Rindler-AdS black hole metric\footnote{We set the AdS radius as $1$.}, which is given as
\begin{align}
	ds^2\,&=\,-(r^2-1)dt^2+\frac1{r^2-1} dr^2+r^2d\mathbf x^2\, ,\label{Rindler-AdS metric}\\
	d\mathbf x^2&=\rho^{-2}(d\rho^2+d\mathbf x_\perp^2)\, ,
\end{align}
where $\mathbf x$ is a coordinate on hyperbolic space $\mathbb H^{d-1}$, and the Hawking temperature, which would be the temperature of the boundary theory, is $1/2\pi$. The horizon of the black hole is on $r=1$, and in order to give a regular incoming boundary condition near the horizon, we need a coordinate system that is regular in the vicinity of the horizon. Eddington-Finkelstein coordinates $(v,r,\mathbf x)$ are appropriate for this purpose, and the metric \eqref{Rindler-AdS metric} can be written as
\begin{equation}
	ds^2\,=\,-(r^2-1)dv^2+2dv dr+r^2d\mathbf x^2\, , \label{EF coordinate metric}
\end{equation}
where $v=r_*+t$ with the tortoise coordinate $r_*=\log\left(\frac{r-1}{r+1}\right)^{1/2}$. 

On the Eddington-Finkelstein coordinates, bulk fields that we consider are spin-$0,1,\ell\geq2$ fields:
\begin{equation}
\begin{split}	
	h(v,r,\mathbf x)\ &:\ \text{Scalar field}\,,\\
	h_{\mu}(v,r,\mathbf x)\ &:\ \text{Vector field}\,,\\ 
	h_{\mu_1\dots\mu_\ell}(v,r,\mathbf x)\ &:\ \text{Symmetric traceless rank-$\ell$ tensor fields ($\ell\geq2$)}\, .
\end{split}
\end{equation}
We can make an ansatz for  field solutions by separation of the variables $v,r,\mathbf x$ as
\begin{equation}
	h_{\mu_1\dots \mu_\ell}(v,r,\mathbf x)=e^{-i\omega v} h_{\mu_1\dots \mu_\ell}(r)f(\mathbf x)\, , \label{field ansatz}
\end{equation}
where $f(\mathbf x)$ is an eigenfunction of the Laplacian operator $\Box_{\mathbb H}$ on $(d-1)$-dimensional hyperbolic space:
\begin{equation}
	\Box_{\mathbb H}f(\mathbf x)=\lambda_{\mathbb H}f(\mathbf x)\, , \label{hyperbolic eigenvalue}
\end{equation}
with an eigenvalue $\lambda_{\mathbb H}$ \cite{Ohya:2016gto, Haehl:2019eae}. For large geodesic distance $\mathbf d\gg 1$,\footnote{The geodesic distance between two points $\mathbf{d}:=\mathbf d(\mathbf x,\mathbf x')$ on hyperbolic space $\mathbb H^{d-1}$ is given by \begin{equation*}\cosh\mathbf d(\mathbf x_1,\mathbf x_2)=\frac{\rho_1^2+\rho_2^2+({\mathbf x}_{\perp1}-{\mathbf x}_{\perp2})^2}{2\rho_1\rho_2}\,.\end{equation*}} the eigenfunction $f(\mathbf x)$ is proportional to  $e^{-\mu\mathbf d}$, where $\lambda_{\mathbb H}=\mu(\mu-d+2)$ \cite{Ahn:2019rnq,Haehl:2019eae}. Since we are interested in asymptotic solutions of the fields near the horizon $r=1$, we now expand the functions depending on $r$ in the field solutions \eqref{field ansatz} as
\begin{equation}
    h_{\mu_1\dots \mu_\ell}(r)=\sum_{j=0}^\infty h_{\mu_1\dots \mu_\ell}^{(j)} (r-1)^{j} \, . \label{asymptotic expansion}
\end{equation}

After plugging these expansions into the EOMs and comparing them for each $(r-1)^j$ order, we can get expansion coefficients and determine the asymptotic field solutions. However, specifying the asymptotic solutions is not our interest. Rather, we want to figure out situations where the asymptotic solutions depend on extra free parameters and the solutions are not uniquely determined. In the following sections, we explicitly figure out the asymptotic solutions of the EOMs of fields with spin-$0,1,\ell\geq2$ and seek special points where the field solutions still have the extra free parameters.

\subsection{Near-horizon analysis of scalar fields (\texorpdfstring{$\ell=0$}{l=0})}\label{subsec:nhas}

Near-horizon analysis of scalar fields on the Rindler-AdS black hole was studied in detail by \cite{Ahn:2020baf}. By reviewing it, we investigate the tower structure of special points of the scalar fields for generalizations to higher-spin fields.

The EOM of a free scalar field $h(v,r,\mathbf x)$ with mass $m^2=\Delta(\Delta-d)$ on the Rindler-AdS black hole metric \eqref{EF coordinate metric} becomes
\begin{equation}
\begin{split}
	0=&(\nabla_\mu\nabla^\mu-\Delta(\Delta-d))h(v,r,\mathbf x)\\
	=&(r^2-1)h''(r)+((d+1)r-(d-1)r^{-1}-2i\omega)h'(r)\\
	&+(r^{-2}\lambda_{\mathbb H}-i\omega(d-1)r^{-1}-\Delta(\Delta-d))h(r)\,, \label{h EOM}
\end{split}
\end{equation}
where $\nabla_\mu$ is the covariant derivative, prime is the derivative with respect to $r$, and we inserted the field solution ansatz \eqref{field ansatz}. We drop the overall factor $e^{-i\omega v}f(\mathbf x)$ henceforth. The EOM for the scalar field becomes a second order ordinary differential equation (ODE) of the function $h(r)$. To determine the asymptotic expansion of the field solution near the horizon $r=1$, we apply the series expansion ansatz \eqref{asymptotic expansion} with $\ell=0$ into the EOM \eqref{h EOM} and get
\begin{equation}
\begin{split}
	&\left[(\lambda_{\mathbb H}-\Delta(\Delta-d)-(d-1)i\omega)h^{(0)}+(2-2i\omega)h^{(1)}\right](r-1)^0\\
	+&\left[(-2\lambda_{\mathbb H}+(d-1)i\omega)h^{(0)}+(\lambda_{\mathbb H}-\Delta(\Delta-d)-(d-1)i\omega+2d)h^{(1)}\right.\\
	&\left.\qquad+(4-2i\omega)h^{(2)}\right](r-1)^1+\big[\cdots\big](r-1)^2+\dots=0\, .	\label{expanded h EOM}
\end{split}
\end{equation} 

In general, we can specify the field solution near the horizon by requiring the vanishment of each expansion coefficient with respect to $(r-1)^j$. More explicitly, from the coefficient of $(r-1)^0$ in \eqref{expanded h EOM}, we have
\begin{equation}
	(\lambda_{\mathbb H}-\Delta(\Delta-d)-(d-1)i\omega)h^{(0)}+(2-2i\omega)h^{(1)}=0\, . \label{h EOM level1}
\end{equation}
Thus, except for the solutions at special points, $h^{(1)}$ can be written in terms of $h^{(0)}$. From the coefficient of $(r-1)^1$ in \eqref{expanded h EOM}, we have
\begin{equation}
	(-2\lambda_{\mathbb H}+(d-1)i\omega)h^{(0)}+(\lambda_{\mathbb H}-\Delta(\Delta-d)-(d-1)i\omega+2d)h^{(1)}+(4-2i\omega)h^{(2)}=0\,, \label{h EOM level2}
\end{equation}
and $h^{(2)}$ can be written in terms of $h^{(0)}$ and $h^{(1)}$. Moreover, we can express $h^{(2)}$ in terms of $h^{(0)}$ by using \eqref{h EOM level1}. Likewise, the higher-order coefficients  $h^{(j)}$ of the field solution can be induced from the higher-order series expansion in \eqref{expanded h EOM} and can be written in terms of the coefficients on the lower orders. By iterating this process, one can determine all the coefficients $h^{(j)}$ only by $h^{(0)}$. Finally, $h^{(0)}$ can be fixed by the field normalization so that the asymptotic field solution $h(v,r,\mathbf x)$ can be uniquely determined near the horizon without any free parameters.

However, as we alerted earlier, there are special points where the above chain of determining the expansion coefficients still contains extra free parameters. For instance, the coefficients of $h^{(0)}$ and $h^{(1)}$ in \eqref{h EOM level1} vanish for the following values of $(\omega, \lambda_{\mathbb H})$%
\begin{equation}
	i\omega=1\,,\quad \lambda_{\mathbb H}=(\Delta-1)(\Delta-d+1) \, . \label{h special points level1}
\end{equation} 
Thus, $h^{(1)}$ cannot be determined from $h^{(0)}$, and the asymptotic field solution still depends on $h^{(1)}$ even after fixing the normalization. Therefore, at the points \eqref{h special points level1}, which are determined from \eqref{expanded h EOM} at the leading order $(r-1)^0$, the asymptotic field solution depends on the extra free parameter, and we call such points \eqref{h special points level1} the leading special points.
Even though we luckily avoid the leading special points \eqref{h special points level1}, the same thing can occur at the sub-leading order $(r-1)^1$. To see it, we use the fact that $h^{(1)}$ can be written in terms of $h^{(0)}$ except for the points \eqref{h special points level1}. By using this relation, the equation \eqref{h EOM level2} at the sub-leading order $(r-1)^1$ is written by a linear combination of $h^{(0)}$ and $h^{(2)}$. Just as the same case for the special points at the leading order, we can also seek special points where the coefficients of $h^{(0)}$ and $h^{(2)}$ vanish simultaneously. The coefficient of $h^{(2)}$ easily determines the value of $i\omega$ at the special points: $i\omega=2$. Since the coefficient of $h^{(0)}$ becomes quadratic in $\lambda_{\mathbb H}$ or quartic in $\mu$, there are four special points in terms of $\mu$:
\begin{equation}
	i\omega=2\,,\quad \lambda_{\mathbb H}=\Delta(\Delta-d+2),\ (\Delta-2)(\Delta-d)\,, \label{h special points level2}
\end{equation}
where $\lambda_{\mathbb H}=\mu(\mu-d+2)$.
By iterating these processes, one can find further special points from the higher-orders $(r-1)^j$ in \eqref{expanded h EOM}. 

We comment on conditions of $h^{(0)}$ at special points. Due to (\ref{h EOM level1}), any regular solutions (\ref{field ansatz}) with $i\omega=1$ satisfy 
\begin{align}
\left(\Box_{\mathbb H}-(\Delta-1)(\Delta-d+1)\right)h^{(0)}f(\mathbf x)=0,\label{nts2}
\end{align}
which has a solution $h^{(0)}=0$.
The field solution with $h^{(0)}=0$ can be written in terms of $h^{(1)}$ and has no extra free parameter after fixing the normalization. Thus, we do not regard $h^{(0)}=0$ as a special point. In other words, $h^{(0)}$ at the leading special points should satisfy (\ref{nts2}) and the following condition:
\begin{align}
h^{(0)} \text{ is not the zero function}.\label{nts}
\end{align}
We call conditions like (\ref{nts}) ``nonzero conditions" for special points.

Instead of (\ref{nts}), it is also convenient to consider the following equation:
\begin{align}
\left(\Box_{\mathbb H}-(\Delta-1)(\Delta-d+1)\right)g^{(0)}=\mathcal{N}\delta(\mathbf{x},\mathbf{x}'),\label{eop}
\end{align}
where $\mathcal{N}$ is a normalization constant, and $\delta(\mathbf{x},\mathbf{x}')$ is a delta function on $\mathbb H^{d-1}$. Due to the delta function, $g^{(0)}$ is not the zero function, and we can ignore the delta function if $\mathbf{x}\ne\mathbf{x}'$. Therefore, $g^{(0)}$ with $\mathbf{x}\ne\mathbf{x}'$ can capture the behaviors (\ref{nts2}) and (\ref{nts}) at the leading special points. In Section \ref{sec:si}, we will explicitly construct integrals that are related to $g^{(0)}$.

Now we briefly review a method that one can systematically determine special points for all orders \cite{Blake:2019otz}. To do so, we collect the coefficients of the field solution as a column vector $\vec h=(h^{(0)},h^{(1)},h^{(2)},\dots)^T$. The equations coming from the vanishment of each order of $(r-1)^{j}$ in \eqref{expanded h EOM} can now be expressed as a matrix multiplication $M^{(\ell=0)}\cdot \vec h=0$ with 
\begin{equation}
	M^{(\ell=0)}=\begin{pmatrix}
		M_{00} & M_{01} & 0 & 0 & 0 &\cdots\\
		M_{10} & M_{11} & M_{12} & 0 & 0 &\cdots\\
		M_{20} & M_{21} & M_{22} & M_{23} & 0 &\cdots\\
		\vdots & \vdots & \vdots & \vdots & \vdots & 
	\end{pmatrix}\, ,
\end{equation}
where 
\begin{align*}
	&M_{00}=\lambda_{\mathbb H}-\Delta(\Delta-d)-(d-1)i\omega\, ,\\
	&M_{10}=	-2\lambda_{\mathbb H}+(d-1)i\omega\, ,\\
	&M_{11}=\lambda_{\mathbb H}-\Delta(\Delta-d)-(d-1)i\omega+2d\, ,\\
	&\qquad\qquad\vdots\\
	&M_{n(n+1)} =2(n+1-i\omega)\,\qquad (n\geq0)\, .
\end{align*}
The matrix $M^{(\ell=0)}$ has a form of a lower-triangular matrix with the additional off-diagonal components $M_{n(n+1)}$, where $n\geq0$.
In terms of the components $M_{ij}$, the equations for each row of $M^{(\ell=0)}\cdot \vec h=0$ become
\begin{align}
	M_{00}h^{(0)}+M_{01}h^{(1)}=0\ ,&\label{h leading M equations}\\
	\left(\prod\limits_{j=0}^{n-1}-M_{j(j+1)}\right)^{-1}\det \mathcal M^{(n)}h^{(0)}+M_{n(n+1)}h^{(n+1)}=0\ ,&\qquad (n\geq1) \label{h higher M equations}
\end{align}
where $\mathcal M^{(n)}$ is a square sub-matrix of $M$ with the components $\mathcal M^{(n)}_{ij}=M_{ij}$, where $i,j=0,1,\dots, n$. The equations \eqref{h higher M equations} can be derived by using the expressions of $h^{(n)}$ in terms of the lower order coefficients, where we assume $M_{j(j+1)}\neq0$ for $j=0,1,\dots, n-1$. Thus, we can systematically figure out special points from (\ref{h leading M equations}) and (\ref{h higher M equations}) by imposing the following conditions:
\begin{equation}
	M_{n(n+1)}=0\ ,\qquad \det\mathcal M^{(n)}=0\ ,\qquad (n\geq0)\,,
\end{equation}
which give the results
\begin{equation}
	i\omega=n+1\ ,\quad \lambda_{\mathbb H}=(\Delta-n+2q-1)(\Delta-n+2q-d+1)\ ,\quad (n\geq0,\ q=0,1,\cdots,n)\,.\label{sptower}
\end{equation}
At large $\mathbf d$, after we restore the temperature $T$, these special points represent exponential behaviors as
\begin{equation}
    e^{-2\pi T (n+1)  v}e^{-\mu\mathbf d},\qquad \mu\ =\ \begin{cases}(\Delta-n+2q-1)\\-(\Delta-n+2q-d+1)\end{cases}\,,\label{tssp}
\end{equation}
where $n\geq0,\ q=0,1,\cdots,n$.

\subsection{Near-horizon analysis of vector fields (\texorpdfstring{$\ell=1$}{l=1})}\label{subsec:nhav}

Now we move on to special points of vector fields. The EOMs of a free vector field $h_\mu(v,r,\mathbf x)$ with respect to $h_v, h_r$ and the Lorenz condition $\nabla^\mu h_\mu=0$ are written as \cite{Costa:2014kfa}
\begin{align}
\begin{split}
	&(\nabla_\mu\nabla^\mu-\Delta(\Delta-d)+1) h_v(v,r,\mathbf x)\\
	&=(r^2-1)h_v''(r)+((d-1)r-(d-1)r^{-1}-2i\omega)h_v'(r)\\
	&\quad+(r^{-2}\lambda_{\mathbb H}-i\omega(d-1)r^{-1}-(d-1)-\Delta(\Delta-d))h_v(r)-2i\omega r h_r(r)=0\, ,\label{hvEOM}\\
\end{split}
\\
\begin{split}
	&(\nabla_\mu\nabla^\mu-\Delta(\Delta-d)+1) h_r(v,r,\mathbf x)\\
	&=(r^2-1)h_r''(r)+\left((d+1)r-(d-1)r^{-1}-2i\omega+2r\right)h_r'(r)\\
	&\quad+\left(r^{-2}\lambda_{\mathbb H}-i\omega (d-1)r^{-1}-\Delta(\Delta-d)+(d-1)r^{-2}+2\right)h_r(r)\\
	&\quad-(d-1)r^{-2}h_v(r)-2(re^{-i\omega v}f(\mathbf x))^{-1}\nabla^i h_{i}(v,r,\mathbf x)=0\, ,\label{hrEOM}
\end{split}
\\
\begin{split}
	&\nabla^\mu h_\mu(v,r,\mathbf x)\\
	&= h_v'(r)+(r^2-1) h'_r(r)+(d-1)r^{-1}h_v(r) \\
	&\quad+\left((d+1)r-(d-1)r^{-1}-i\omega\right) h_r(r)+(e^{-i\omega v}f(\mathbf x))^{-1}\nabla^i h_{i}(v,r,\mathbf x)=0\, .\label{lorenz}
\end{split}
\end{align}
The EOM of the vector component $h_v$ \eqref{hvEOM} is a coupled ODE of $h_v(r)$ and $h_r(r)$. The EOM of $h_r$ \eqref{hrEOM} is a coupled ODE of $h_v(r)$, $h_r(r)$, and $\nabla^i h_{i}(v,r,\mathbf x)$.  By combining the EOM of $h_r$ \eqref{hrEOM} and the Lorenz condition \eqref{lorenz}, one can reduce the three equations (\ref{hvEOM}), (\ref{hrEOM}), and (\ref{lorenz}) into two coupled ODEs, which contains only the two field components $h_v(r)$ and $h_r(r)$:
\begin{align}
\begin{split}
	&(r^2-1)h_v''(r)+((d-1)r-(d-1)r^{-1}-2i\omega)h_v'(r)\\
	&\quad+(r^{-2}\lambda_{\mathbb H}-i\omega(d-1)r^{-1}-(d-1)-\Delta(\Delta-d))h_v(r)-2i\omega r h_r(r)=0\,, \label{hEOM1}
\end{split}
\\[2mm]
\begin{split}
	&(r^2-1)h_r''(r)+((d+5)r-(d+1)r^{-1}-2i\omega)h_r'(r)+2r^{-1}h_v'(r)+(d-1)r^{-2}h_v(r)\\
	&+(-i\omega(d+1)r^{-1}-\Delta(\Delta-d)+2d+4+(\lambda_{\mathbb H}-d+1)r^{-2})h_r(r)=0\,. \label{hEOM2}
\end{split}
\end{align}
By inserting the field expansion of each component around the horizon like \eqref{asymptotic expansion}, at the leading order, we have
\begin{equation}
	(\lambda_{\mathbb H}-\Delta(\Delta-d)-(d-1)-i\omega(d-1))h_v^{(0)}-2i\omega h_r^{(0)}-2i\omega h_v^{(1)}=0\,, \label{vector constraint1}
\end{equation}
\begin{equation}
	(\lambda_{\mathbb H}-\Delta(\Delta-d)+d+5-i\omega(d+1))h_r^{(0)}+(d-1) h_v^{(0)}+(4-2i\omega )h_r^{(1)}+2h_v^{(1)}=0\,. \label{vector constraint2}
\end{equation}
In usual cases, these two constraints (\ref{vector constraint1}) and (\ref{vector constraint2}) determine $h_v^{(1)}$ and $\ h_r^{(1)}$ in terms of $h_v^{(0)}$ and $\ h_r^{(0)}$. However, at special points
\begin{equation}
	i\omega=0\ ,\qquad \lambda_{\mathbb H}=(\Delta-1)(\Delta-d+1)\,,\label{lpsvf}
\end{equation}
the constraint \eqref{vector constraint1} becomes trivial, and the other constraint \eqref{vector constraint2} becomes
\begin{equation}
	(d-1)h_v^{(0)}+(2d+4)h_r^{(0)}+2h_v^{(1)}+4h_r^{(1)}=0\, ,  \label{vector possibility 1}
\end{equation}
and $h_v^{(1)}$ or $h_r^{(1)}$ cannot be determined only  by $h_v^{(0)}$ and $\ h_r^{(0)}$. At other special points
\begin{equation}
	i\omega=2\ ,\qquad \lambda_{\mathbb H}=(\Delta-1)(\Delta-d+1)\,, \label{inferior point of vector}
\end{equation}
 two constraints (\ref{vector constraint1}) and (\ref{vector constraint2}) boil down to one constraint
\begin{equation}
	(d-1)h_v^{(0)}+2h_r^{(0)}+2h_v^{(1)}=0\,. \label{vector possibility 2}
\end{equation}
In this case, $h_r^{(1)}$ cannot be determined  by $h_v^{(0)},\ h_r^{(0)}$ and becomes a free parameter.
We can repetitively do this procedure in higher-order and get additional special points.

We can also decouple the two coupled ODEs (\ref{hEOM1}) and (\ref{hEOM2}) so that we can apply the technique for scalar fields.
By defining two linear combinations  
\begin{align}
\begin{split}
	H_0(r)&\equiv h_v(r)+(r^2-r)h_r(r)\,,\\ 
	H_1(r)&\equiv h_v(r)+(r^2+r)h_r(r)\,,\label{HEOM}
\end{split}	
\end{align}
we can decouple \eqref{hEOM1} and \eqref{hEOM2} into two decoupled ODEs 
\begin{align}
\begin{split}
	&(r^2-1)H_0''(r)+((d+1)r-(d-1)r^{-1}-2(i\omega+1))H_0'(r)\\
	&+(r^{-2}\lambda_{\mathbb H}-(i\omega+1)(d-1)r^{-1}-\Delta(\Delta-d))H_0(r)=0\, , \label{AEOM}
\end{split}
\\[2mm]
\begin{split}
	&(r^2-1)H_1''(r)+((d+1)r-(d-1)r^{-1}-2(i\omega-1))H_1'(r)\\
	&+(r^{-2}\lambda_{\mathbb H}-(i\omega-1)(d-1)r^{-1}-\Delta(\Delta-d))H_1(r)=0\, . \label{BEOM}	
\end{split}
\end{align}
Now we can repeat the same process for the scalar field case to find all the special points of these decoupled ODEs. Note that the ODE of $H_1(r)$ is the same as that of $H_0(r)$ replaced by $i\omega\rightarrow i\omega-2$. The special points of \eqref{AEOM} turn out to be
\begin{equation}
	i\omega=n\ ,\quad \lambda_{\mathbb H}=(\Delta-n+2q-1)(\Delta-n+2q-d+1)\ ,\quad (n\geq0,\ q=0,1,\cdots,n)\, . \label{spin-1 special points}
\end{equation}
The special points of \eqref{BEOM} are equal to the points of \eqref{AEOM} replaced by $i\omega\rightarrow i\omega-2$: 
\begin{equation}
	i\omega=n+2\ ,\quad \lambda_{\mathbb H}=(\Delta-n+2q-1)(\Delta-n+2q-d+1)\ ,\quad (n\geq0,\ q=0,1,\cdots,n)\, . \label{spin-1 sub special points}
\end{equation}

By using a fact that (\ref{AEOM}) and (\ref{BEOM}) are equal to the ODE of scalar fields \eqref{h EOM} where $i\omega$ is replaced by $i\omega+1$ or $i\omega-1$, one can directly obtain the results (\ref{spin-1 special points}) and (\ref{spin-1 sub special points}) from (\ref{sptower}). Note that the points \eqref{spin-1 sub special points} are a subset of the special points \eqref{spin-1 special points} and, the leading special points in \eqref{spin-1 sub special points} were already revealed in \eqref{inferior point of vector}. Thus, the overall special point structure of the vector fields ($\ell=1$) associated with the components $v$ and $r$ is given by \eqref{spin-1 special points}.  Our analysis here does not give special points associated with $\nabla^i h_{i}=0$. At the large $\mathbf d$ limit, these special points represent exponential behaviors 
\begin{equation}
    e^{-2\pi T n  v}e^{-\mu\mathbf d},\qquad \mu\ =\ \begin{cases}(\Delta-n+2q-1)\\-(\Delta-n+2q-d+1)\end{cases}\,,
\end{equation}
where $n\geq0,\ q=0,1,\cdots,n$.

Our results (\ref{spin-1 special points}) and (\ref{spin-1 sub special points}) show that special points in decoupled ODEs of different linear combinations are different in general. As another example, special points of gauge-invariant variables constructed from a massless gauge field on the Rindler-AdS black hole were studied in \cite{Ahn:2020baf}. The points  derived in \cite{Ahn:2020baf} are not (\ref{spin-1 special points}), but a subset of (\ref{spin-1 special points}).

To give an understanding of the connection between the approaches from coupled differential equations and from decoupled differential equations, let us expand $H_0(r)$ and $H_1(r)$ (\ref{HEOM}) as
\begin{align}
H_0(r)&=h_v^{(0)}+[h_v^{(1)}+h_r^{(0)}](r-1)+\dots\,,\\
H_1(r)&=[h_v^{(0)}+2h_r^{(0)}]+[h_v^{(1)}+3h_r^{(0)}+2h_r^{(1)}](r-1)+\dots\,.\label{H1EOM}
\end{align}
At the points \eqref{lpsvf}, the coefficients of $(r-1)^0$ and $(r-1)^1$ in $H_0(r)$ are still independent. As the coefficient of $(r-1)^1$ cannot be determined from the one of $(r-1)^0$, \eqref{lpsvf} are special points of $H_0(r)$. At the same point (\ref{lpsvf}), the asymptotic expansion of $H_1(r)$ \eqref{H1EOM} becomes
\begin{align}
H_1(r)&=[h_v^{(0)}+2h_r^{(0)}]+(-d/2+1/2)[h_v^{(0)}+2h_r^{(0)}](r-1)+\dots\,,
\end{align}
where we used \eqref{vector possibility 1}. The second coefficient depends on the first one, and it is consistent with the fact that \eqref{lpsvf} are not special points of $H_1(r)$.

\subsection{Near-horizon analysis of symmetric traceless rank-\texorpdfstring{$\ell$}{l} tensor fields}

The EOMs and divergence-free conditions of symmetric traceless rank-$\ell$ tensor fields \cite{Bouatta:2004kk,Costa:2014kfa} on the Rindler-AdS black hole metric \eqref{Rindler-AdS metric} are
\begin{align}
\begin{split}
	&(\nabla_\mu\nabla^\mu-\Delta(\Delta-d)+\ell) h_{v\dots v}(v,r,\mathbf x)\\
	&=(r^2-1)h_{v\dots v}''+\left((d+1-2\ell)r-(d-1)r^{-1}-2i\omega\right)h_{v\dots v}'\\
	&\quad+\left(r^{-2}\lambda_{\mathbb H}-i\omega(d-1)r^{-1} -\ell (d-1) -\Delta(\Delta-d)\right)h_{v\dots v}(v,r,\mathbf x)\\
	&\quad-\left(2i\ell r\omega +2\ell(\ell-1)r^2\right) h_{v\dots vr}(v,r,\mathbf x)-\ell(\ell-1)r^2(r^2-1)h_{v\dots vrr}(v,r,\mathbf x)=0\, ,\label{EFEOMl}
\end{split}
\\
\begin{split}
	&(\nabla_\mu\nabla^\mu-\Delta(\Delta-d)+1) h_{v\dots vr}(v,r,\mathbf x)\\
	&=(r^2-1)h_{v\dots vr}''+\left((d+1)r-(d-1)r^{-1}-2i\omega-2(\ell-2)r\right)h_{v\dots vr}'\\
	&\quad+\left(r^{-2}\lambda_{\mathbb H}-i\omega (d-1)r^{-1}-\ell(d-1)-\Delta(\Delta-d)+d+1+r^{-2}(d-1)\right)h_{v\dots vr}\\
	&\quad-\left(2i\omega(\ell-1)r+2(\ell-1)(\ell-3)r^2\right)h_{v\dots vrr}-(\ell-1)(\ell-2)r^2(r^2-1)h_{v\dots vrrr}\\
	&\quad-(d-1)r^{-2}h_{v\dots v}-2(re^{-i\omega v}f(\mathbf x))^{-1}\nabla^i h_{v\dots vi}(v,r,\mathbf x)=0\,,
\end{split}
\\&\qquad\qquad\vdots\nonumber\\
\begin{split}
	&\nabla^\mu h_{v\dots v\mu}(v,r,\mathbf x)\\
	&= h'_{v\dots v}(r)+(r^2-1) h'_{v\dots vr}(r)+(d-1)r^{-1}h_{v\dots v}(r)\\
	&\quad +\left((d+1)r -(d-1)r^{-1}-i\omega \right)h_{v\dots vr}(r)+(e^{-i\omega v}f(\mathbf x))^{-1}\nabla^i h_{v\dots vi}(v,r,\mathbf x)=0\,.
\end{split}
\\&\qquad\qquad\vdots\nonumber
\end{align}
We can obtain a few special points $(n=0,1,\dots)$ starting from the leading special points as we did in the vector field case even though the equations are complicatedly coupled differential equations. The leading special points of symmetric traceless rank-$\ell$ tensor fields were first determined by this method in \cite{Kim:2021hqy}: 
\begin{equation}
	i\omega=1-\ell\ ,\qquad \lambda_{\mathbb H}=(\Delta-1)(\Delta-d+1)\, . \label{spin-l leading special points}
\end{equation}

However, it is laborious to obtain the whole special point structure by this method. Instead, one can decouple these coupled differential equations, as we did in the previous subsection. 
Then we can obtain the whole special point structure by applying the matrix method for the scalar field case with $\ell=0$ to each decoupled ODE.
For higher-spin fields, it is more difficult to decouple the EOMs because we need at least $\ell+1$ coupled differential equations (the EOMs of $h_{v\dots v},\,h_{v\dots vr},\,\dots,\, h_{r\dots r}$). Here, the divergence-free conditions are used to make the EOMs into forms without  $\nabla^{i}h_{\mu_1\dots\mu_{\ell-1}i}$.  For the case with $\ell=2$, we find three specific linear combinations\footnote{We redundantly use the symbol $H$ for linear combinations of fields with different spin $\ell$. Depending on the spin of fields, there are $\ell+1$ linear combinations $H_0,H_1,\dots,H_\ell$}. 
\begin{align}
\begin{split}
H_{0}(r)&\equiv h_{vv}(r)+2r(r-1)h_{vr}(r)+r^2(r-1)^2h_{rr}(r)\,,\\
H_{1}(r)&\equiv h_{vv}(r)+2r^2h_{vr}(r)+r^2(r-1)(r+1)h_{rr}(r)\,,\\
H_{2}(r)&\equiv h_{vv}(r)+2r(r+1)h_{vr}(r)+r^2(r+1)^2h_{rr}(r)	\,,
\end{split}
\end{align}
which decouple the three EOMs of $h_{vv}$, $h_{vr}$, and $h_{rr}$ with the divergence-free  conditions into
\begin{align}
\begin{split}
	&(r^2-1)H_0''(r)+((d+1)r-(d-1)r^{-1}-2(i\omega+2))H_0'(r)\\
	&+(r^{-2}\lambda_{\mathbb H}-(i\omega+2)(d-1)r^{-1}-\Delta(\Delta-d))H_0(r)=0\, , \label{H0 EOM}
\end{split}
\\[2mm]
\begin{split}
	&(r^2-1)H_1''(r)+((d+1)r-(d-1)r^{-1}-2i\omega)H_1'(r)\\
	&+(r^{-2}\lambda_{\mathbb H}-i\omega(d-1)r^{-1}-\Delta(\Delta-d))H_1(r)=0\, , \label{H1 EOM}
\end{split}
\\[2mm]
\begin{split}
	&(r^2-1)H_2''(r)+((d+1)r-(d-1)r^{-1}-2(i\omega-2))H_2'(r)\\
	&+(r^{-2}\lambda_{\mathbb H}-(i\omega-2)(d-1)r^{-1}-\Delta(\Delta-d))H_2(r)=0\, . \label{H2 EOM}
\end{split}
\end{align}
The ODEs of $H_1(r)$ and $H_2(r)$ can be obtained by replacing $i\omega\rightarrow i\omega-2,\, i\omega-4$ in \eqref{H0 EOM}. After repeating the same technique that determines special points of the ODE in the scalar field ($\ell=0$) case, special points of \eqref{H0 EOM} turn out to be
\begin{equation}
	i\omega=n-1\ ,\quad \lambda_{\mathbb H}=(\Delta-n+2q-1)(\Delta-n+2q-d+1)\ ,\quad (n\geq0,\ q=0,1,\cdots,n)\,, \label{spin-2 special points}
\end{equation}
and special points of (\ref{H1 EOM}) and (\ref{H2 EOM}) are equal to the special points \eqref{spin-2 special points} replaced by $i\omega\rightarrow i\omega-2$ or $i\omega\rightarrow i\omega-4$, which are subsets of \eqref{spin-2 special points}.

We also confirm the special point structure of symmetric traceless rank-$3$ tensor fields. The decoupled ODEs obtained from the four EOMs and the divergence-free conditions in terms of four specific linear combinations
\begin{align*}
H_{0}(r)&\equiv h_{vvv}(r)+3r(r-1)h_{vvr}(r)+3r^2(r-1)^2h_{vrr}(r)+r^3(r-1)^3h_{rrr}(r)\,,\\
H_{1}(r)&\equiv h_{vvv}(r)+r(3r-1)h_{vvr}(r)+r^2(3r^2-2r-1)h_{vrr}(r)+r^3(r-1)^2(r+1)h_{rrr}(r)\,,\\
H_{2}(r)&\equiv h_{vvv}(r)+r(3r+1)h_{vvr}(r)+r^2(3r^2+2r-1)h_{vrr}(r)	+r^3(r-1)(r+1)^2h_{rrr}(r)\,,\\
H_{3}(r)&\equiv h_{vvv}(r)+3r(r+1)h_{vvr}(r)+3r^2(r+1)^2h_{vrr}(r)+r^3(r+1)^3h_{rrr}(r)	\,,
\end{align*}
are given by 
\begin{align}
\begin{split}
	&(r^2-1)H_0''(r)+((d+1)r-(d-1)r^{-1}-2(i\omega+3))H_0'(r)\\
	&+(r^{-2}\lambda_{\mathbb H}-(i\omega+3)(d-1)r^{-1}-\Delta(\Delta-d))H_0(r)=0\, , \label{spin3 ODE}
\end{split}
\end{align}
and the other three ODEs of $H_1(r)$, $H_2(r)$, and $H_3(r)$ are equal to \eqref{spin3 ODE} replaced by $i\omega\rightarrow i\omega-2,\,i\omega-4,i\omega-6$, respectively.  Special points of the symmetric traceless rank-$3$ tensor fields turn out to be
\begin{equation}
	i\omega=n-2\ ,\quad \lambda_{\mathbb H}=(\Delta-n+2q-1)(\Delta-n+2q-d+1)\ ,\quad (n\geq0,\ q=0,1,\cdots,n)\, . \label{spin-3 special points}
\end{equation}

Though we do not complete the general procedure to decouple the differential equations for $\ell>3$, we can infer from our results with $\ell=0,1,2,3$ that the special point structure of symmetric traceless rank-$\ell$ tensor fields is expected to be
\begin{equation}
	i\omega=n-\ell+1\ ,\quad \lambda_{\mathbb H}=(\Delta-n+2q-1)(\Delta-n+2q-d+1)\ ,\quad (n\geq0,\ q=0,1,\cdots,n)\, , \label{spin-l special points}
\end{equation}
where the leading special points ($n=0$) are revealed in \cite{Kim:2021hqy} for general integer $\ell$. Exponential behaviors at the large $\mathbf d$ limit corresponding to these inferred results are
\begin{equation}
    e^{2\pi T(\ell-n-1) v}e^{-\mu\mathbf d},\qquad \mu\ =\ \begin{cases}(\Delta-n+2q-1)\\-(\Delta-n+2q-d+1)\end{cases}\,, \label{spin-l exp}
\end{equation}
where $n\geq0,\ q=0,1,\cdots,n$.

\section{Integrals of scalar propagators}\label{sec:si}
Assuming that special points with $i\omega=2\pi (n+1)T$ exist, we construct integrals of scalar bulk-to-bulk propagators over black hole horizons that capture behaviors at the special points in the near-horizon analysis, where $n\in\{0\}\cup\mathbb{N}$. First, we study the construction on the Rindler-AdS black hole as a solvable example. Next, we consider the construction in a large class of static black holes.

\subsection{Integrals for the leading special points on the Rindler-AdS black hole}\label{subsec:lpss}
First, we construct integrals for the leading special points with $i\omega=1$ on the Rindler-AdS black hole. Scalar bulk-to-bulk propagator $G_\Delta\left(\xi\right)$ between two bulk points $Y$ and $Y'$ on the Rindler-AdS black hole is given by \cite{Costa:2014kfa, Ahn:2020csv}\footnote{For simplicity, the normalization of (\ref{bbprorbh}) is different from the one in \cite{Costa:2014kfa, Ahn:2020csv}.}
\begin{align}
G_\Delta\left(\xi\right)=&\left(\frac{\xi}{2}\right)^\Delta\,_2F_1\left(\frac{\Delta}{2}, \frac{\Delta+1}{2}, \Delta+1-\frac{d}{2}; \xi^2\right),\label{bbprorbh}\\
\xi:=&\frac{(1+UV)(1+U'V')}{2(UV'+VU')+(1-UV)(1-U'V')\cosh \mathbf{d}}\,,\label{rbhxi}
\end{align}
where we use the Kruskal-Szekeres coordinates
\begin{align}
ds^2\,=\,-\frac{4dUdV}{(1+UV)^2}+\left(\frac{1-UV}{1+UV}\right)^2d\mathbf x^2\,.
\end{align}
When $\Delta$ satisfies the unitarity bound $\Delta\ge d/2-1$, the three parameters of the hypergeometric function are non-negative. 
Up to a delta function, $G_\Delta\left(\xi\right)$ is a solution of the EOM
\begin{align}
\left(\nabla_\mu\nabla^\mu-\Delta(\Delta-d)\right)G_\Delta(\xi)=0.\label{seom}
\end{align}
The singular behavior of the hypergeometric function at $\xi^2=1$ in (\ref{bbprorbh}) is due to the delta function. To avoid this singularity, we impose the following constraint:
\begin{align}
UV\le0, \;\;\; U'V'\le0, \;\;\; UV'\ge0, \;\;\; VU'\ge0, \;\;\; \mathbf{d}>0.\label{constraint}
\end{align}
Under this constraint, we obtain $|\xi|<1$. To be more explicit, we choose $Y$ for the right region of the Penrose diagram and $Y'$ for the left region as follows:
\begin{align}
U=-e^{r_*-t}, \;\;\; V=e^{r_*+t}, \;\;\; U'=e^{r_*'-t'}, \;\;\; V'=-e^{r_*'+t'}.\label{UVrelation}
\end{align}

Consider an integral of $G_\Delta\left(\xi\right)$ over the horizon $V'=0$ as
\begin{align}
\int_{0}^{\infty}dU' G_\Delta(\xi)|_{V'=0}.\label{sint1}
\end{align}
By introducing a new variable $U'':=VU'$, we see that (\ref{sint1}) is proportional to $V^{-1}$ as
\begin{align}
\int_{0}^{\infty}dU' G_\Delta(\xi)|_{V'=0}=\frac{1}{V}\int_0^{\infty}dU''G_\Delta\left(\frac{1}{\cosh \mathbf{d}r+(r+1)U''}\right)\,.
\end{align}
From the near-horizon analysis in Subsection \ref{subsec:nhas}, we can evaluate the $\mathbf{d}$-dependence of (\ref{sint1}) at $r=1$ without performing the integration as follows. Since $G_\Delta\left(\xi\right)$ is a solution of the EOM, (\ref{sint1}) is also a solution if the integral converges. Therefore, (\ref{sint1}) at $r=1$, which is proportional to $V^{-1}=e^{-2\pi Tv}=e^{-v}$, satisfies 
\begin{align}
\big(\Box_{\mathbb H}- (\Delta-1)(\Delta-d+1)\big)\int_{0}^{\infty}dU' G_\Delta(\xi)|_{V'=0, r=1}=0,\label{selps}
\end{align}
which is valid if $\mathbf{d}\ne0$.

Let us confirm that $\int_{0}^{\infty}dU' G_\Delta(\xi)|_{V'=0, r=1}$ is a nonzero function.
One can explicitly calculate $\int_{0}^{\infty}dU' G_\Delta(\xi)|_{V'=0, r=1}$ \cite{Kim:2021hqy} as
\begin{align}
&\int_{0}^{\infty}dU' G_\Delta(\xi)|_{V'=0, r=1}=\frac{1}{V}\int_0^{\infty}dU''G_\Delta\left(\frac{1}{\cosh \mathbf{d}+2U''}\right)\notag\\
=&\frac{1}{2^{\Delta+1}V}\int_0^{1/\cosh \mathbf{d}}d\xi \xi^{\Delta-2}\,_2F_1\left(\frac{\Delta}{2}, \frac{\Delta+1}{2}, \Delta+1-\frac{d}{2}; \xi^2\right)\notag\\
=&\frac{1}{4(\Delta-1)V}e^{-(\Delta-1)\mathbf{d}}\,_2F_1\left(\Delta-1, \frac{d}{2}-1, \Delta+1-\frac{d}{2}; e^{-2\mathbf{d}}\right),\label{sslsp}
\end{align}
where we assume $\Delta>1$ for convergence of the integral. In fact, (\ref{sslsp}) is known as a solution of (\ref{selps}) up to a delta function \cite{Cornalba:2006xm, Cornalba:2007fs}. Since the eigenvalue $\Delta(\Delta-d)$ in (\ref{seom}) is invariant under $\Delta\leftrightarrow d-\Delta$, $\int_{0}^{\infty}dU' G_{d-\Delta}(\xi)|_{V'=0}$ with $d-\Delta>1$ is another solution for the leading special points. These two solutions have different asymptotic behaviors at large $\mathbf{d}$ as
\begin{align}
\int_{0}^{\infty}dU' G_{\Delta}(\xi)|_{V'=0, r=1}\simeq& \frac{1}{4(\Delta-1)V}e^{-(\Delta-1)\mathbf{d}} \;\;\; (\mathbf{d}\gg1), \\
\int_{0}^{\infty}dU' G_{d-\Delta}(\xi)|_{V'=0, r=1}\simeq& \frac{1}{4(d-\Delta-1)V}e^{-(d-\Delta-1)\mathbf{d}} \;\;\; (\mathbf{d}\gg1),\label{ablps}
\end{align}
which agree with (\ref{tssp}) at $n=0$.

When $\mathbf{d}=0$, the hypergeometric function in (\ref{sslsp}) diverges, and we need to add the delta function to the right hand side of (\ref{selps}). In particular, $\int_{0}^{\infty}dU' G_\Delta(\xi)|_{V'=0, r=1}$ is a solution of (\ref{eop}). The singular behavior of $G_\Delta(\xi)$ at $\mathbf{d}=0$ is related to the fact that $\int_{0}^{\infty}dU' G_\Delta(\xi)|_{V'=0, r=1}$ is a nonzero function as explained around (\ref{eop}).

Summarizing the above, the integral (\ref{sint1}) is a nonzero function proportional to $V^{-1}=e^{-2\pi Tv}=e^{-v}$ and satisfies (\ref{selps}) if $\mathbf{d}\ne0$. Therefore, (\ref{sint1}) captures the behavior at the leading special points, \textit{i.e.}, (\ref{nts2}), (\ref{nts}), and $i\omega=1$. In this paper, we assume that integrals we construct satisfy nonzero conditions for special points like (\ref{nts}) due to the singular behaviors of propagators.

\subsection{Integrals for the sub-leading special points on the Rindler-AdS black hole}\label{sec:sls}
Next, we construct integrals for the sub-leading special points with $i\omega=n+1$, where $n\in\mathbb{N}$. Our strategy is to differentiate $\int_{0}^{\infty}dU' G_\Delta(\xi)$ with respect to $V'$. As a first example, let us consider
\begin{align}
\int_{0}^{\infty}dU' \partial_{V'}G_\Delta(\xi)|_{V'=0}=\frac{1}{V}\int_{0}^{\infty}dU'' \frac{\partial\xi}{\partial V'}\Big|_{V'=0}\partial_{\xi}G_\Delta(\xi)|_{\xi=\frac{1}{\cosh \mathbf{d}r+(r+1)U''}},\label{sint2}
\end{align}
which is a solution of EOM up to a delta function if the integral converges. We decompose $\frac{\partial\xi}{\partial V'}\Big|_{V'=0}$ into two parts for later convenience:
\begin{align}
\frac{\partial\xi}{\partial V'}\Big|_{V'=0}=&U'\frac{\partial \xi}{\partial (U'V')}\Big|_{V'=0}+U\frac{\partial \xi}{\partial (UV')}\Big|_{V'=0},\label{Vderivative}\\
U'\frac{\partial \xi}{\partial (U'V')}\Big|_{V'=0}=&\frac{1}{V}U''\left[\frac{1}{\cosh \mathbf{d}r+(r+1)U''}+\frac{r\cosh \mathbf{d}}{\left(\cosh \mathbf{d}r+(r+1)U''\right)^2}\right],\\
U\frac{\partial \xi}{\partial (UV')}\Big|_{V'=0}=&\frac{r-1}{V}\frac{1}{\left(\cosh \mathbf{d}r+(r+1)U''\right)^2},\label{UV'derivative}
\end{align}
where we use $U=-\frac{r-1}{V(r+1)}$. Since (\ref{Vderivative}) is proportional to $V^{-1}$, (\ref{sint2}) is proportional to $V^{-2}$. Therefore, assuming that the singular behavior of the propagator is related to nonzero conditions for special points, (\ref{sint2}) captures the behavior at the sub-leading special points with $i\omega=2$. More generally, one can argue that $\int_{0}^{\infty}dU' (\partial_{V'})^nG_\Delta(\xi)|_{V'=0}$ are proportional to $V^{-(n+1)}$ and capture the behaviors at special points with $i\omega=n+1$.

Let us evaluate $\int_{0}^{\infty}dU' \partial_{V'}G_\Delta(\xi)|_{V'=0, r=1}$ at large $\mathbf{d}$. By using an approximation $G_\Delta\left(\xi\right)\simeq\left(\xi/2\right)^\Delta$ at small $\xi$, we obtain
\begin{align}
\int_{0}^{\infty}dU' \partial_{V'}G_\Delta(\xi)|_{V'=0, r=1}\simeq\frac{1}{8V^2(\Delta-2)}e^{-(\Delta-2)\mathbf{d}} \;\;\; (\mathbf{d}\gg1),\label{ssslsp1}
\end{align}
where we assume $\Delta>2$ for the convergence. Since (\ref{UV'derivative}) becomes zero at $r=1$, (\ref{UV'derivative}) does not contribute to (\ref{ssslsp1}). At the sub-leading order, (\ref{UV'derivative}) contributes to (\ref{sint2}) as
\begin{align}
&\frac{1}{V}\int_{0}^{\infty}dU''U\frac{\partial \xi}{\partial (UV')}\Big|_{V'=0, r\to1}\partial_{\xi}G_\Delta(\xi)|_{\xi=\frac{1}{\cosh \mathbf{d}+2U''}}\notag\\
\simeq&\frac{r-1}{2V^2}e^{-\Delta\mathbf{d}} \;\;\; (\mathbf{d}\gg1), \label{ssslsp2}
\end{align}
where we assume $\Delta>0$, and the exponential behavior is different from the one of (\ref{ssslsp1}). These two exponential behaviors (\ref{ssslsp1}) and (\ref{ssslsp2}) agree with (\ref{tssp}) at $n=1$.

We emphasize again the difference between the near-horizon analysis in Section \ref{sec:nha} and the analysis of $\int_{0}^{\infty}dU' (\partial_{V'})^nG_\Delta(\xi)|_{V'=0}$ in this section. In the near-horizon analysis, we searched conditions that $h^{(0)}$ can be a nonzero extra free parameter and obtained the special points (\ref{sptower}) at $i\omega=n+1$. In this section, on the other hand, we showed that $\int_{0}^{\infty}dU' (\partial_{V'})^nG_\Delta(\xi)|_{V'=0}$ are proportional to $V^{-i\omega}=e^{-i\omega v}$ with $i\omega=n+1$. If $\int_{0}^{\infty}dU' (\partial_{V'})^nG_\Delta(\xi)|_{V'=0, r=1}$ are nonzero functions, the integrals capture the behaviors at the special points with $i\omega=n+1$.

\subsection{Integrals for special points in static black holes}\label{subsec:sbhs}
Finally, we give a construction method for special points in a static black hole
\begin{align}
ds^2=A(UV)dUdV+B(UV)d\mathbf{x}_M^2,\label{sbhm}
\end{align}
where $A(UV)$ and $B(UV)$ are functions of $UV=f(r)$ for the static metric, $d\mathbf{x}_M^2$ is a metric on a manifold $M$. We define $\mathbf{d}_M$ as the spatial distance on $M$ and $r_H$ as the radius of a horizon such that $UV(r_H)=f(r_H)=0$ in the Kruskal-Szekeres coordinates and $g^{rr}(r_H)=0$ in the incoming Eddington-Finkelstein coordinates, where we assume the unique horizon radius. Let us consider a scalar propagator $G$ between two bulk points $Y$ and $Y'$ in the black hole (\ref{sbhm}), which is a solution of a linear second-order differential EOM up to a delta function, with the following assumption:
\begin{itemize}
\item The scalar propagator $G$ is invariant under the isometry of (\ref{sbhm}). Specifically, $G$ is a function of $UV$, $U'V'$, $UV'$, $VU'$, and $\mathbf{d}_M$.\footnote{The metric (\ref{sbhm}) is invariant under $U\to\lambda U, V\to V/\lambda$. Correspondingly, the scalar propagator $G(UV, U'V', UV', VU', \mathbf{d}_M)$ is invariant under $U\to\lambda U, V\to V/\lambda, U'\to\lambda U', V'\to V'/\lambda$.} We express this property as $G(UV, U'V', UV', VU', \mathbf{d}_M)$.
\end{itemize}
Generally, it is difficult to find an analytic expression of $G(UV, U'V', UV', VU', \mathbf{d}_M)$. The propagator (\ref{bbprorbh}) is an example that satisfies the assumption.

Just like (\ref{sint1}), consider an integral over the horizon $V'=0$ as
\begin{align}
\int_{0}^{\infty}dU' G(UV, U'V', UV', VU', \mathbf{d}_M)|_{V'=0},\label{slpsgbh}
\end{align}
where we assume convergence of the integral and regularity at $r=r_H$, except where $G(UV, U'V', UV', VU', \mathbf{d}_M)$ diverges due to the delta function. By introducing a new variable $U'':=VU'$, we obtain
\begin{align}
\int_{0}^{\infty}dU' G(UV, U'V', UV', VU', \mathbf{d}_M)|_{V'=0}=\frac{1}{V}\int_{0}^{\infty}dU'' G(f(r), 0, 0, U'', \mathbf{d}_M),
\end{align}
which is proportional to $V^{-1}$.

As shown in the example in Subsection \ref{subsec:nhas}, recent holographic studies of the pole-skipping phenomenon reveal towers of pole-skipping points in Green's functions and special points in the near-horizon analysis at imaginary Matsubara frequencies \cite{Grozdanov:2019uhi,Blake:2019otz}.  Thus, we assume that special points of scalar fields exist at $i\omega=2\pi (n+1)T$. Since (\ref{slpsgbh}) is proportional to $V^{-1}=e^{-2\pi T v}$, (\ref{slpsgbh}) satisfies an equation like (\ref{nts2}) at $i\omega=2\pi T$. As explained in Subsection \ref{subsec:lpss}, we assume that (\ref{slpsgbh}) also satisfies a nonzero condition like (\ref{nts}) due to the singular behavior of the scalar propagator. Thus, we argue that (\ref{slpsgbh}) captures the behavior at the leading special points with $i\omega=2\pi T$ in the black hole (\ref{sbhm}).

Next, consider an integral 
\begin{align}
\int_{0}^{\infty}dU' \partial_{V'}G(UV, U'V', UV', VU', \mathbf{d}_M)|_{V'=0},\label{sslpsgbh}
\end{align}
which is a generalization of (\ref{sint2}). By using $UV=f(r)$ and $U''=VU'$, we obtain
\begin{align}
&\int_{0}^{\infty}dU' \partial_{V'}G(UV, U'V', UV', VU', \mathbf{d}_M)|_{V'=0}\notag\\
=&\frac{1}{V^2}\int_{0}^{\infty}dU''\Big[U''\frac{\partial}{\partial U'V'}G(f(r), U'V', UV', U'', \mathbf{d}_M)\notag\\
&+f(r)\frac{\partial}{\partial UV'}G(f(r), U'V', UV', U'', \mathbf{d}_M)\Big]\Big|_{V'=0},
\end{align}
which is proportional to $V^{-2}$ and capture the behavior at the sub-leading special points with $i\omega=4\pi T$. 
Similarly, one can construct $\int_{0}^{\infty}dU' \partial^n_{V'}G(UV, U'V', UV', VU', \mathbf{d}_M)|_{V'=0}$ for the sub-leading special points with $i\omega=2\pi (n+1)T$.

\section{Integrals of spin-\texorpdfstring{$\ell$}{l} propagators}\label{sec:sli}
We construct integrals for special points in the near-horizon analysis of symmetric traceless rank-$\ell$ tensor fields. We mainly analyze vector fields with $\ell=1$ and consider a generalization to arbitrary integer spin $\ell$.

\subsection{Integrals for special points of vector fields}
Let us consider a vector propagator $G_{\mu}^{\;\;\mu'}$ between two bulk points $Y$ and $Y'$ in the black hole (\ref{sbhm}). By introducing a polarization vector $Z'$ of $Y'$, we contract the index $\mu'$ of $G_{\mu}^{\;\;\mu'}$ and express the propagator as $G_\mu(Y, Y', Z')$, which is a first-order polynomial of $Z'$. We suppose that $G_\mu(Y, Y', Z')$ is a solution of a linear second-order differential EOM in the black hole (\ref{sbhm}) up to a delta function. For an example of $G_\mu(Y, Y', Z')$, see the AdS propagator in \cite{Costa:2014kfa}.

To satisfy $Y'^\mu Z'_\mu=0$ at $V'=0$, we choose the polarization vector $Z'$ as
\begin{align}
Z'_{U}=0, \;\;\; Z'_{V}\ne0, \;\;\; Z'_{\mathbf{x}_M}=0.\label{lpv}
\end{align}
Due to the index $\mu$, tensor structures of $G_\mu(Y, Y', Z')$ are different from ones of scalar propagators. We assume the following structure of $G_\mu(Y, Y', Z')$ at $V'=0$:
\begin{align}
G_\mu(Y, Y', Z')=&Y_{\mu}\, g^1_1(Y, Y')(VZ'_V)+Y'_{\mu}\, g^1_2(Y, Y')(VZ'_V)+Z'_{\mu}\, g^1_3(Y, Y'),\label{vp}
\end{align}
where functions $g^1_i(Y, Y')$ are invariant under the isometry, and we use $Y'^\mu Z'_\mu=0$.

With the above preparation, consider the $V$-dependence of 
\begin{align}
\int_{0}^{\infty}dU' G_\mu(Y, Y', Z')|_{V'=0}.
\end{align}
By using (\ref{lpv}) and the similar analysis in Subsection \ref{subsec:sbhs} with $U'':=VU'$, we obtain
\begin{align}
&\int_{0}^{\infty}dU' G_U(Y, Y', Z')|_{V'=0}\propto V^1, \;\;\; \int_{0}^{\infty}dU' G_V(Y, Y', Z')|_{V'=0}\propto V^{-1},\notag\\
&\int_{0}^{\infty}dU' G_{\mathbf{x}_M}(Y, Y', Z')|_{V'=0}\propto V^{0}.
\end{align}
Let us transform the integrals with $\mu=U, V, \mathbf{x}_M$ in the Kruskal-Szekeres coordinates to the ones with $\mu=r, v, \mathbf{x}_M$ in the Eddington-Finkelstein coordinates. With a transformation rule
\begin{align}
\int_{0}^{\infty}dU' G_v(Y, Y', Z')|_{V'=0}&=\frac{\partial V}{\partial v}\int_{0}^{\infty}dU' G_V(Y, Y', Z')|_{V'=0},\\
\int_{0}^{\infty}dU' G_U(Y, Y', Z')|_{V'=0}&=\left(\frac{\partial U}{\partial V}\right)^{-1}\int_{0}^{\infty}dU' G_V(Y, Y', Z')|_{V'=0}\notag\\
&+\left(\frac{\partial U}{\partial r}\right)^{-1}\int_{0}^{\infty}dU' G_r(Y, Y', Z')|_{V'=0},\\
\frac{\partial V}{\partial v}\propto V, \;\;\; &\frac{\partial U}{\partial V}\propto V^{-2}, \;\;\; \frac{\partial U}{\partial r}\propto V^{-1},
\end{align}
the final result is given by
\begin{align}
\int_{0}^{\infty}dU' G_\mu(Y, Y', Z')|_{V'=0}\propto V^{0} \;\;\; (\mu=r, v, \mathbf{x}_M).\label{vlpsgbh}
\end{align}
As a further generalization, one can obtain
\begin{align}
\int_{0}^{\infty}dU' \partial^n_{V'}G_\mu(Y, Y', Z')|_{V'=0}\propto V^{-n} \;\;\; (\mu=r, v, \mathbf{x}_M). \label{vslpsgbh}
\end{align}

In Subsection \ref{subsec:nhav}, we showed that the special points of vector fields on the Rindler-AdS black hole exist at $i\omega=2\pi nT$, where $n\in\{0\}\cup\mathbb{N}$. Suppose that these special points also exist in the EOMs of vector fields in the black hole (\ref{sbhm}). Since the integrals (\ref{vslpsgbh}) are also solutions of the EOMs that are proportional to $V^{-n}=e^{-2\pi nTv}$, we argue that the integrals (\ref{vslpsgbh}) capture the behaviors at special points with $i\omega=2\pi nT$. 

\subsection{Integrals for special points of spin-\texorpdfstring{$\ell$}{l} fields}
Consider propagators $G_{\mu_1\dots \mu_\ell}(Y, Y', Z')$ of symmetric traceless rank-$\ell$ tensor fields, which are solutions of the EOMs up to delta functions and $\ell$th-order polynomials of $Z'$. We assume the following property of $G_{\mu_1\dots \mu_\ell}(Y, Y', Z')$ at $V'=0$:
\begin{enumerate}[i]
\item Tensor structures of $G_{\mu_1\dots \mu_\ell}(Y, Y', Z')$ are determined from $Y_\mu$,  $Y'_\mu$, $Z'_\mu$, and the metric $g_{\mu_i\mu_j}$. The propagators $G_{\mu_1\dots \mu_\ell}(Y, Y', Z')$ at $V'=0$ can be expressed by these tensor structures, $VZ'_V$, and functions $g^\ell_i(Y, Y')$ invariant under the isometry of (\ref{sbhm}).\label{A-1}
\end{enumerate}
For example, $G_{\mu_1\mu_2}(Y, Y', Z')$ with $Y'^\mu Z'_\mu=0$ under the assumption can be expressed as
\begin{align}
G_{\mu_1\mu_2}(Y, Y', Z')=&Y_{\mu_1}Y_{\mu_2}g^2_1(Y, Y')(VZ'_V)^2+Y'_{\mu_1}Y'_{\mu_2}g^2_2(Y, Y')(VZ'_V)^2\notag\\
+&Y_{\{\mu_1}Y'_{\mu_2\}}g^2_3(Y, Y')(VZ'_V)^2+g_{\mu_1\mu_2}g^2_4(Y, Y')(VZ'_V)^2\notag\\
+&Y_{\{\mu_1}Z'_{\mu_2\}}g^2_5(Y, Y')(VZ'_V)+Y'_{\{\mu_1}Z'_{\mu_2\}}g^2_6(Y, Y')(VZ'_V)\notag\\
+&Z'_{\mu_1}Z'_{\mu_2}g^2_7(Y, Y'),
\end{align}
where $g^2_i(Y, Y')$ are invariant under the isometry, and $Y_{\{\mu_1}Y'_{\mu_2\}}:=(Y_{\mu_1}Y'_{\mu_2}+Y_{\mu_2}Y'_{\mu_1})/2$. The same analysis as in the previous subsection yields
\begin{align}
\int_{0}^{\infty}dU' \partial^n_{V'}G_{\mu_1\mu_2}(Y, Y', Z')|_{V'=0}\propto V^{1-n} \;\;\; (\mu_i=r, v, \mathbf{x}_M),
\end{align}
and a generalization for arbitrary $\ell$ is
\begin{align}
\int_{0}^{\infty}dU' \partial^n_{V'}G_{\mu_1\dots\mu_\ell}(Y, Y', Z')|_{V'=0}\propto V^{\ell-1-n} \;\;\; (\mu_i=r, v, \mathbf{x}_M).\label{slfslpsgbh}
\end{align}

In the study of the pole-skipping phenomena, the leading special points of gravitons have been found at $i\omega=-2\pi T$ \cite{Blake:2018leo}. On the Rindler-AdS black hole, the leading special points of spin-$\ell$ fields were found at $i\omega=2\pi (1-\ell)T$ \cite{Kim:2021hqy}, and the towers of special points (\ref{spin-l exp}) at $i\omega=2\pi (n+1-\ell)T$ seem to be present from the analysis in Section \ref{sec:nha}. If these special points of spin-$\ell$ fields in the black hole (\ref{sbhm}) also exist at $i\omega=2\pi (n+1-\ell)T$, the integrals (\ref{slfslpsgbh}) capture the behaviors at these special points.

For the convenience of the reader, we summarize the assumptions of our argument.
\begin{itemize}
\item The propagators $G_{\mu_1\dots \mu_\ell}(Y, Y', Z')$ of symmetric traceless rank-$\ell$ tensor fields, which are $\ell$th-order polynomials of the polarization vector (\ref{lpv}), are solutions of linear second-order differential EOMs in the static black hole (\ref{sbhm}) up to delta functions and satisfy the property \ref{A-1}.
\item Except where $G_{\mu_1\dots \mu_\ell}(Y, Y', Z')$ diverge due to the delta functions, the integrals $\int_{0}^{\infty}dU' \partial^n_{V'}G_{\mu_1\dots\mu_\ell}(Y, Y', Z')|_{V'=0}$, where $n\in\{0\}\cup\mathbb{N}$, are well-defined: they are regular at $r=r_h$ and converge.
\end{itemize}
From these assumptions, one can argue (\ref{slfslpsgbh}). In addition to them, we assume the following:
\begin{itemize}
\item Special points of the spin-$\ell$ fields exist at $i\omega=2\pi (n+1-\ell)T$, where $T$ is the Hawking temperature of the black hole (\ref{sbhm}). Specifically, there are  independent regular solutions of the EOMs that are proportional to $V^{\ell-1-n}$.
\item The integrals (\ref{slfslpsgbh}) with $\mathbf{x}_M\ne\mathbf{x}'_M$ satisfy nonzero conditions for the special points like (\ref{nts}).
\end{itemize}
Then, we argue that the integrals (\ref{slfslpsgbh}) capture the behaviors at the special points. As discussed in Subsection \ref{subsec:lpss}, the singular behaviors of the propagators seem to be related to 
the assumption that the integrals (\ref{slfslpsgbh}) satisfy the nonzero conditions.

\section{Interpretation of the integrals for special points}\label{sec:interpret}
We discuss the interpretation of the integrals we constructed for special points by using four-point amplitudes with the exchange of a spin-$\ell$ field. We see that late time behaviors of the amplitudes are related to the integrals due to bulk-to-bulk propagators in the amplitudes.

\begin{figure}
\centering
     {\includegraphics[width=3cm]{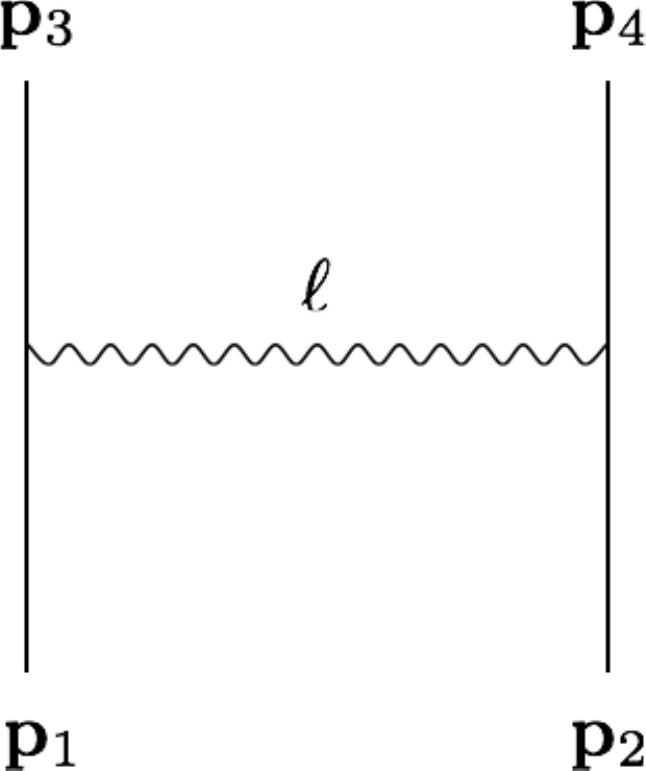}}
 \caption{Scalar four-point tree-level diagram with the exchange of a spin-$\ell$ field in AdS$_{d+1}$, where $\mathbf{p}_i$ are points on the AdS boundary, straight lines represent external wave functions $\phi_i$ of scalar fields, and a wavy line represents a spin-$\ell$ bulk-to-bulk propagator $\mathbf{\Pi}_{a_1\dots a_\ell}^{\;\;\;\;\;\;\;\;\;\;a'_1\dots a'_\ell}$.}\label{ted}
\end{figure} 

\subsection{Integrals for the leading special points}
We start reviewing a scalar four-point tree-level diagram with the exchange of a spin-$\ell$ field in AdS$_{d+1}$ as in Figure \ref{ted} \cite{Cornalba:2006xk}. To represent AdS$_{d+1}$, we use $\mathbf{X}=(x^+, x^-, x)$\footnote{Our coordinates $\mathbf{X}$ correspond to $\mathbf{x}$ in \cite{Cornalba:2006xk}.} in the embedding space $\mathbb{R}^{1, 1} \times \mathbb{R}^{1, d-1}$, where $x^+$ and $x^-$ are light cone coordinates on the Minkowski space $\mathbb{R}^{1, 1}$, and $x$ is a point in the Minkowski space $\mathbb{R}^{1, d-1}$. With these embedding coordinates, AdS$_{d+1}$ with the AdS radius $L=1$ can be represented by
\begin{align}
-x^+x^-+x^2=-1.
\end{align}
With a three-point derivative coupling between the scalar and spin-$\ell$ fields, the amplitude of the tree-level diagram in Figure \ref{ted} is given by \cite{Cornalba:2006xk}
\begin{align}
&\int_{\text{AdS}_{d+1}}\widetilde{d\mathbf{X}}\left[\phi_3\left(\prod_{i=1}^\ell\partial^{a_i}\right)\phi_1\right] \mathbf{h}_{a_1\dots a_\ell}(\mathbf{X}),\\
&\mathbf{h}_{a_1\dots a_\ell}(\mathbf{X}):=\int_{\text{AdS}_{d+1}}\widetilde{d\mathbf{X}'}\mathbf{\Pi}_{a_1\dots a_\ell}^{\;\;\;\;\;\;\;\;\;\;a'_1\dots a'_\ell}(\mathbf{X},\mathbf{X}')\mathbf{T}_{a'_1\dots a'_\ell}(\mathbf{X}'),\label{cpslf}\\
&\mathbf{T}_{a'_1\dots a'_\ell}(\mathbf{X}'):=\phi_4\left(\prod_{i=1}^\ell\partial_{a'_i}\right)\phi_2,
\end{align}
where $\widetilde{d\mathbf{X}}$ is an integral measure in AdS$_{d+1}$, $\phi_i$ are external wave functions of the scalar fields, $\mathbf{\Pi}_{a_1\dots a_\ell}^{\;\;\;\;\;\;\;\;\;\;a'_1\dots a'_\ell}(\mathbf{X},\mathbf{X}')$ is a bulk-to-bulk propagator of the spin-$\ell$ field, and we ignore an unimportant normalization constant.

To make a shock wave, the authors of \cite{Cornalba:2006xk} chose the wave functions $\phi_2$ and $\phi_4$ so that the source $\mathbf{T}_{a'_1\dots a'_\ell}(\mathbf{X}')$ is localized along $x'^-=0$ and $a'_i=-$. Moreover, we choose $\mathbf{T}_{a'_1\dots a'_\ell}(\mathbf{X}')$ localized at $y$ in transverse hyperbolic space $\mathbb{H}^{d-1}$ as
\begin{align}
\mathbf{T}_{a'_1\dots a'_\ell}(\mathbf{X}')\propto\left(\prod_{i=1}^\ell\delta_{a'_i}^-\right)\delta(x'^-)\delta(x',y),\label{sps}
\end{align}
where $\delta(x',y)$ is a delta function in $\mathbb{H}^{d-1}$. Due to these delta functions, the $(d+1)$-dimensional integral in (\ref{cpslf}) with (\ref{sps}) becomes a one-dimensional integral in the $x'^+$-direction.

\begin{figure}
\centering
     {\includegraphics[width=9cm]{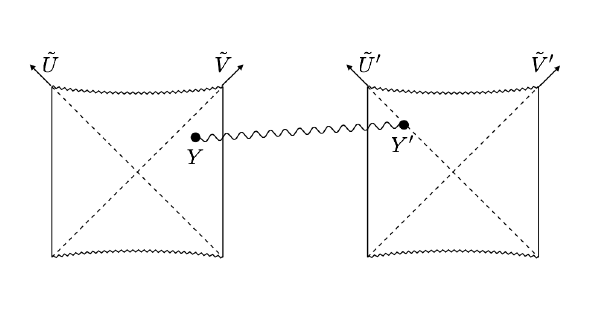}}
 \caption{Two bulk points $Y$ and $Y'$ in Penrose diagrams, where we introduce new conformal coordinates $\tilde{U}(U)$ and $\tilde{V}(V)$. The Penrose diagrams on the left and right are for points $\mathbf{x}_M$ and $\mathbf{x}'_M$, respectively. A wavy line represents  the propagator $G_{\mu_1\dots \mu_\ell}^{\;\;\;\;\;\;\;\;\;\;\;\mu'_1\dots \mu'_\ell}(Y,Y')$.}\label{bpipd}
\end{figure}

Our integrals (\ref{slfslpsgbh}) for the leading special points with $n=0$ can be interpreted as a generalization of (\ref{cpslf}) with (\ref{sps}) in the black hole (\ref{sbhm}). The light cone coordinates $x^+$ and $x^-$ correspond to $U$ and $V$,\footnote{More precisely, $U$ and $V$ in the Rindler-AdS black hole correspond to $u^+/2$ and $u^-/2$ in \cite{Cornalba:2006xk}. When $u^+=0$ or $u^-=0$, the two coordinates are equal such as $x^+=u^+$ and $x^-=u^-$.} and the propagator $\mathbf{\Pi}_{a_1\dots a_\ell}^{\;\;\;\;\;\;\;\;\;\;a'_1\dots a'_\ell}(\mathbf{X},\mathbf{X}')$ corresponds to $G_{\mu_1\dots \mu_\ell}^{\;\;\;\;\;\;\;\;\;\;\;\mu'_1\dots \mu'_\ell}(Y,Y')$. The localization of (\ref{sps}) along $a'_i=-$ corresponds to the polarization vector (\ref{lpv}), and the $x'^+$-integral at $x'^-=0$ corresponds to the $U'$-integral at $V'=0$. When $\mathbf{X}$ approaches to the point where $\mathbf{T}_{a'_1\dots a'_\ell}(\mathbf{X}')$ is localized, $\mathbf{h}_{a_1\dots a_\ell}(\mathbf{X})$ becomes singular. To avoid the singular behavior of (\ref{slfslpsgbh}), we choose two bulk points $Y$ and $Y'$ from different regions of Penrose diagrams as shown in Figure \ref{bpipd}, which is related to (\ref{UVrelation}) for the Rindler-AdS black hole.

When we consider a graviton propagator with $\ell=2$, (\ref{cpslf}) with (\ref{sps}) is metric deformation for a shock wave geometry \cite{Cornalba:2006xk}. The Lyapunov exponent $\lambda_L$ and the butterfly velocity $v_B$ in holographic models can be calculated from the metric deformation \cite{Blake:2016wvh,Roberts:2016wdl}. Since (\ref{slfslpsgbh}) with $n=0$ is a generalization of (\ref{cpslf}), we would also obtain $\lambda_L$ and $v_B$ from (\ref{slfslpsgbh}). In fact, the Lyapunov exponent $\lambda_L=2\pi T$ is deduced from the $V$-dependence $V\propto e^{2\pi T t}$ of (\ref{slfslpsgbh}), and the butterfly velocity $v_B$ is obtained from the $\mathbf{d}_M$-dependence of (\ref{slfslpsgbh}) at $r=r_H$ and large $\mathbf{d}_M$, which measures propagation of the source in the spatial direction. Since (\ref{slfslpsgbh}) are the integrals for special points, the $\mathbf{d}_M$-dependence can be computed from equations at the special points like (\ref{selps}). This explains why the leading special points of metric perturbation in the near-horizon analysis, which corresponds to the  energy-momentum tensor in the boundary theories, can detect $\lambda_L$ and $v_B$ in the holographic models.

\subsection{Integrals for the sub-leading special points}
The classical profile (\ref{cpslf}) of the spin-$\ell$ field with the localized source (\ref{sps}) at the horizon leads to the leading exponential behavior of the tree-level amplitude with respect to time.
One might expect that the integrals for the sub-leading special points would be related to sub-leading corrections of the tree-level amplitude. This expectation is partially correct, but there are some subtleties. We explain them by using the scalar exchange amplitude in the Rindler-AdS black hole.

\begin{figure}
\centering
     {\includegraphics[width=10cm]{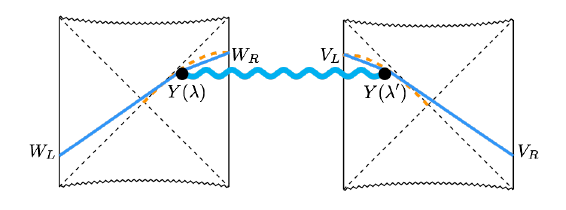}}
 \caption{Half-geodesic Witten diagram in the two-sided Rindler-AdS black hole. The bulk-to-boundary propagators and the bulk-to-bulk propagator are represented by blue straight lines and a blue wavy line, respectively.  Interaction vertices $Y(\lambda)$ and $Y(\lambda')$ are integrated over half-geodesics that are represented by orange dotted curves.}\label{figuregwdrbh}
\end{figure}

As the scalar exchange amplitude, we consider the following amplitude of the half-geodesic Witten diagram as in Figure \ref{figuregwdrbh}: \cite{Kim:2021hqy}
\begin{align}
\begin{split}
\mathcal W^{\mathcal R}_{\Delta,0}:=\int_{0}^{\infty} d\lambda \int_{-\infty}^0 d\lambda' &G_{b\partial}\left(Y(\lambda), W_L; \Delta_W\right)G_{b\partial}\left(Y(\lambda), W_R; \Delta_W\right)\\
\times &G_{\Delta}\left(\xi(\lambda,\lambda')\right) G_{b\partial}\left(Y'(\lambda'), V_L; \Delta_V\right)G_{b\partial}\left(Y'(\lambda'), V_R; \Delta_V\right)\,,\label{gwdrb}
\end{split}
\end{align} 
where $W_L$, $W_R$, $V_L$, and $V_R$ are the AdS boundary points, $G_{b\partial}$ is the scalar bulk-to-boundary propagator, and $G_{\Delta}$ is the scalar bulk-to-bulk propagator (\ref{bbprorbh}). The bulk points $Y(\lambda)$ and $Y'(\lambda')$ are integrated over half geodesics between the AdS boundaries and the centers of diagrams such that 
\begin{align}
Y(\lambda):&\;\;\; U(\lambda)=-e^{-t_W}\tanh \frac{\lambda}{2}\,, \;\;\;\; V(\lambda)=e^{t_W}\tanh \frac{\lambda}{2}\,, \;\;\;\; \mathbf{x}(\lambda)=\mathbf{x}_W,\label{gammaw}\\
Y'(\lambda'):&\;\;\; U'(\lambda')=-e^{-t_V}\tanh \frac{\lambda'}{2}\,, \;\;\; V'(\lambda')=e^{t_V}\tanh \frac{\lambda'}{2}\,, \;\;\; \mathbf{x}'(\lambda')=\mathbf{x}_V\,.
\end{align}
See  \cite{Kim:2021hqy} for more details.

At late times $t_W-t_V=:t_R\gg1$, the bulk-to-boundary propagators can be approximated as constants  \cite{Kim:2021hqy}. Thus, we approximate the half-geodesic Witten diagram (\ref{gwdrb}) in the late time limit as
\begin{align}
\mathcal W^{\mathcal R}_{\Delta,0}\sim\int_{0}^{\infty} d\lambda \int_{-\infty}^0 d\lambda' G_{\Delta}\left(\xi(\lambda,\lambda')\right) \,.\label{apfhgw}
\end{align} 
Since $V'(\lambda')\to0$ in the late time limit, the integral $\int_{-\infty}^0 d\lambda' G_{\Delta}\left(\xi(\lambda,\lambda')\right)$ can be interpreted as the integral (\ref{sint1}). This analysis implies that the late time behavior of the half-geodesic Witten diagram can be detected from the leading special point. In fact, the near-horizon analysis in \cite{Kim:2021hqy} captured the late time behavior.

Let us evaluate sub-leading corrections of (\ref{apfhgw}). Consider a series expansion of $G_{\Delta}\left(\xi\right)$:
\begin{align}
G_{\Delta}\left(\xi\right)=G_{\Delta}\left(\xi\right)|_{V'=0}+\partial_{V'}G_{\Delta}\left(\xi\right)|_{V'=0}+\frac{1}{2}(\partial_{V'})^2G_{\Delta}\left(\xi\right)|_{V'=0}+\dots\,.\label{sep}
\end{align}
The leading behavior of (\ref{apfhgw}) comes from the integral of the first term in (\ref{sep}). The integrals of the other terms give the sub-leading corrections of (\ref{sep}), and these integrals correspond to the solutions $\int_{0}^{\infty}dU' (\partial_{V'})^nG_\Delta(\xi)|_{V'=0}$ constructed in Subsection \ref{sec:sls}. Therefore, the integrals for the sub-leading special points are related to the part of the sub-leading corrections.

Let us discuss problems with this interpretation of the sub-leading special points. First, at the sub-leading order, we need to consider contributions from the bulk-to-boundary propagators, and not all the sub-leading corrections are related to the solutions at the special points. In the late time limit, the four-point amplitude can be approximated from the integral of the bulk-to-bulk propagator. But in general, the four-point amplitude also depends on the bulk-to-boundary propagators. Second, as explained in Subsection \ref{sec:sls}, the convergence of the integral changes. Note that these problems depend on the definition of the amplitude, and it remains to be seen what happens when other amplitudes are used.

\section{Conclusion and discussion}\label{sec:con}

We applied near-horizon analysis to obtain the overall structure of special points by studying the EOMs of bulk free fields with integer spin on the Rindler-AdS black hole. After we reviewed the well-studied special point structure of the scalar fields ($\ell=0)$, we obtained the special point structure of non-zero integer spin fields. For the vector fields ($\ell=1)$, we obtained a few special points from the coupled differential EOMs by searching for the conditions that some series coefficients become extra free parameters, which is the main philosophy of the near-horizon analysis. In principle, one can do this repetitive job to get further special points. Instead, we found specific linear combinations of vector field components that decouple the coupled differential equations into the decoupled ODEs. Thus, we can apply the matrix method to these decoupled ODEs that was used to get the special points of the scalar fields systematically and entirely. In this way, we obtained the overall structure of special points for the massive vector fields, not restricted to the massless vector fields. Not only for the vector fields, but we also found specific linear combinations of the symmetric traceless rank-$\ell$ tensor fields with $\ell=2, 3$ associated with indices $v,r$ such that their EOMs are decoupled. It should be mentioned again that these decouplings are not restricted to massless fields. We expect that this decoupling process is also possible for the fields with $\ell>3$. From our results and predictions, we concluded the overall special point structure of integer spin fields and corresponding exponential behaviors as \eqref{spin-l special points} and \eqref{spin-l exp}. 

We proposed a construction method of integrals that capture behaviors at special points with imaginary Matsubara frequencies $i\omega=2\pi (n+1-\ell)T$, where $\ell$ is spin of bosonic fields, $n$ is a non-negative integer, and $T$ is the Hawking temperature. As seen in (\ref{slfslpsgbh}), our construction by integrating bulk propagators over horizons of static black holes leads to the frequencies $i\omega=2\pi (n+1-\ell)T$. We also discussed the interpretation of our construction in bulk scattering amplitudes on the black hole spacetime and explained that our integrals are generalizations of integrals of graviton propagators for a shock wave geometry, which are related to chaotic properties of theories with the Einstein gravity duals.

Let us discuss some future directions of our work.
We explicitly obtained the special point structure with spin $\ell=0,1,2,3$ and predicted the special points of general spin-$\ell$ fields inductively. This is because it becomes more difficult to decouple the EOMs for higher-spin fields. To do so systematically, we need to understand why such specific linear combinations of field components can decouple the EOMs.  We focused on the field components only with indices $v,r$ in the Eddington-Finkelstein coordinates (\textit{i.e.},  $h_{v\dots v},\,h_{v\dots vr},\,\dots,\,h_{r\dots r}$) because they are expected to give the leading special points \cite{Grozdanov:2019uhi,Blake:2019otz,Natsuume:2019xcy}, and it really does in our case \cite{Kim:2021hqy}. However, the other components of fields (\textit{i.e.},  $h_{v\dots vi},\,h_{v\dots vij},\,\dots,\,h_{ijk\dots}$, where $i,j,k$ are the coordinates on hyperbolic space $\mathbb H^{d-1}$) should be investigated for completeness. One can also generalize our analysis on the Rindler-AdS black hole background \eqref{Rindler-AdS metric} into the more general static black hole background \eqref{sbhm}.

To justify our proposal, it is important to further investigate explicit examples of (\ref{slfslpsgbh}). Specifically, it is worth studying the integrals of propagators of vector or graviton fields. 
In our construction, we used the polarization vector (\ref{lpv}). If we use another polarization vector, we will obtain behaviors with different frequencies than (\ref{slfslpsgbh}), and it would be interesting to examine this difference.

\acknowledgments

We would like to thank Yongjun Ahn for
valuable discussions and comments.
This work was supported by Basic Science Research Program through the National Research Foundation of Korea(NRF) funded by the Ministry of Science, ICT \& Future
Planning (NRF-2021R1A2C1006791) and the GIST Research Institute(GRI) grant funded by the GIST in 2021. 
K. Lee was supported by Basic Science Research Program through the National Research Foundation of Korea(NRF) funded by the Ministry of Education(NRF-2020R1I1A2054376).
M. Nishida was supported by Basic Science Research Program through the National Research Foundation of Korea(NRF) funded by the Ministry of Education(NRF-2020R1I1A1A01072726).


\bibliography{Refs}
\bibliographystyle{JHEP}

\end{document}